\documentclass[11pt]{article}
\pdfoutput=1
\usepackage[dvipsnames]{xcolor}
\usepackage[all]{xy}
\usepackage{amsmath}
\usepackage{amssymb}
\usepackage{braket}
\usepackage{fancyhdr}
\usepackage{hyperref}
\hypersetup{
    unicode=false,          % non-Latin characters in Acrobat’s bookmarks
    pdftoolbar=true,        % show Acrobat’s toolbar?
    pdfmenubar=true,        % show Acrobat’s menu?
    pdffitwindow=false,     % window fit to page when opened
    pdfstartview={FitH},    % fits the width of the page to the window
    pdftitle={My title},    % title
    pdfauthor={Author},     % author
    pdfsubject={Subject},   % subject of the document
    pdfcreator={Creator},   % creator of the document
    pdfproducer={Producer}, % producer of the document
    pdfkeywords={keywords}, % list of keywords
    pdfnewwindow=true,      % links in new window
    colorlinks=false,       % false: boxed links; true: colored links
    linkcolor=red,          % color of internal links
    citecolor=cayn,        % color of links to bibliography
    filecolor=magenta,      % color of file links
    urlcolor=cyan           % color of external links
}
 
\newcommand{\Cwarren}{${\bf F}_{{\color{red}P}{\color{blue}P}{\color{Green}P}}$}
\newcommand{\rP}{{\color{red}P}}
\newcommand{\bP}{{\color{blue}P}}
\newcommand{\gP}{{\color{Green}P}}
\newcommand{\rA}{{\color{red}A}}
\newcommand{\bB}{{\color{blue}B}}
\newcommand{\gC}{{\color{Green}C}}

\newcommand{\gS}{{\color{Green}S}}
\newcommand{\xA}{{\mbox{\footnotesize(1)}}}
\newcommand{\xB}{{\mbox{\footnotesize(2)}}}
\newcommand{\delt}{\delta\,(\,{\mbox{\footnotesize{2}}}\,-\,{\mbox{\footnotesize 1}}\,)}
\newcommand{\deltP}{\delta^{\prime}\,(\,{\mbox{\footnotesize{2}}}\,-\,{\mbox{\footnotesize 1}}\,)}
\newcommand{\defeq}{\mathrel{\mathop:}=}

\def\be{\begin{equation}}
\def\ee{\end{equation}}

\def\frac#1#2{{\textstyle{#1\over#2}}}
\def\Fontamici#1{{$\mathcal{#1}$}}
\def\go#1{{\mbox{{\scriptsize\Fontamici{#1}}}}}

\def\rgboo#1{\pdfliteral{#1 rg #1 RG}}
\def\pdfklink#1#2{%
	\noindent\pdfstartlink user
		{/Subtype /Link
		/Border [ 0 0 0 ]
		/A << /S /URI /URI (#2) >>}{\rgb{1 0 0}{#1}}%
	\pdfendlink}
\def\rgbo#1#2{\rgboo{#1}#2\rgboo{0 0 0}}
\def\rgb#1#2{\mark{#1}\rgbo{#1}{#2}\mark{0 0 0}}

\def\xxxlink#1{\pdfklink{[arXiv:#1]}{http://arXiv.org/abs/#1}}

\def\Gamma{\mathchar"0100}
\def\Delta{\mathchar"0101}
\def\Theta{\mathchar"0102}
\def\Lambda{\mathchar"0103}
\def\Xi{\mathchar"0104}
\def\Pi{\mathchar"0105}
\def\Sigma{\mathchar"0106}
\def\Upsilon{\mathchar"0107}
\def\Phi{\mathchar"0108}
\def\Psi{\mathchar"0109}
\def\Omega{\mathchar"010A}

\begin{document}

\title{{\bf\color{blue} Natural curvature for manifest T-duality}\\[-2in]
{\normalsize September 11, 2013\hfill YITP-SB-13-26}\\[1.8in]}
\date{}
\author{Martin Pol\'a\v cek\footnote{\pdfklink{martin.polacek@stonybrook.edu}{mailto:martin.polacek@stonybrook.edu}},\ \ 
Warren Siegel\footnote{\pdfklink{siegel@insti.physics.sunysb.edu}{mailto:siegel@insti.physics.sunysb.edu},
	\pdfklink{http://insti.physics.sunysb.edu/\~{}siegel/plan.html}{http://insti.physics.sunysb.edu/\%7Esiegel/plan.html}}\\
{\textit{C. N. Yang Institute for Theoretical Physics}}\\ {\textit{State University of New York, Stony Brook, NY 11794-3840}}}

\maketitle

\begin{abstract}

\normalsize

We reformulate the manifestly T-dual description of the massless sector of the closed bosonic string, 
directly from the geometry associated with the (left and right) affine Lie algebra of the coset space Poincar\'e/Lorentz.  
This construction initially doubles not only the (spacetime) coordinates for translations but also those for Lorentz transformations (and their ``dual").  
As a result, the Lorentz connection couples directly to the string (as does the vielbein), rather than being introduced ad hoc to the covariant derivative as previously.
This not only reproduces the old definition of T-dual torsion, but automatically gives a general, covariant definition of T-dual curvature (but still with some undetermined connections).  

\end{abstract}

\newpage

\section{Introduction}

\subsection{Outline}

T-duality invariance can be manifested on all the fields of the massless sector of bosonic strings\nobreak\ \cite{warren}. 
This was based on the treatment of the compactification scalars, for dimensional reduction of $d$ dimensions, as elements of the coset SO(d,d)/SO(d)${}^2$ \cite{Duff}.  
This symmetry was expanded to SO(D,D)/SO(D$-$1,1)${}^2$ for the full $D$ dimensions to include all fields without compactification, where the symmetry is broken spontaneously to the usual SO(D$-$1,1), except when partially restored by dimensional reduction.
(Generalization to GL groups \cite{Mah} was also treated, but turned out not to be convenient for supersymmetry, and will not be considered here.
For relations to later approaches, and extensions beyond what is needed here, see \cite{bow} and references therein.)

We will work on a space with explicit Lorentz coordinates.  Dependence of the (background) vielbein on them is completely fixed (up to gauge) by the coset constraints, as applied by fixing the associated parts of the torsion to take their ``vacuum" values. Moreover, as in \cite{warren}, D-dimensional spacetime will be dualized. To do the stringy generalizations (of oscillator algebras together with the Lorentz algebras), we will need to introduce a new current $\Sigma$ for consistency with the Jacobi identity \cite{warren2}.  (The necessity of this current was first realized in the context of AdS${}_5\times$S${}^5$ \cite{Machiko}.) The usual oscillator Lie algebra will become the extended affine Lie algebra (Lorentz and $\Sigma$ generators included). 

The generalized torsion is constructed from this affine Lie algebra in a general background, which acts as the stringy generalization of covariant derivatives.  Because of the additional currents, the enlarged vielbein that describes this background includes the Lorentz connection, and the enlarged torsion includes also the curvature.  Closure of the algebra implies the orthogonality constraints $E\eta E^{{\bf{T}}}\,=\,\eta$ on the vielbein.  Solving these together with the coset constraints reduces the vielbein to the usual T-dual generalization of the vielbein and Lorentz connection, as well as a new curvature-like field.  There is also an extension of dimensional reduction to the usual D coordinates. At the end we will obtain the same results for the torsion constraints and curvature tensor as previously, but by a much more direct way.   

The rest of this paper is organized as follows:  In the remainder of the Introduction we summarize the general procedure.  In the next section we review the description of fields on general coset spaces, and then apply this to the case of spin for Poincar\'e/Lorentz to give a ``first-quantized" approach to general relativity.  The corresponding affine Lie algebra is described in section 3.  In section 4 we introduce the vielbein and the coset constraints on the torsion, and orthogonality.  The new analysis of Lorentz connections and curvatures is given in section 5, followed by our conclusions.

\subsection{Procedure}

The general procedure (to be applied in detail below for the present example) is thus:
\vspace{-6pt}
\begin{enumerate}
\setlength{\itemsep}{0pt}
\item Begin with a coset space G/H.  By the usual construction (left and right group multiplication) this comes with two Lie algebras for G, one for ``symmetry generators" and one for ``covariant derivatives", represented by derivatives on the group space.
\item Generalize to the affine Lie algebras by making the group coordinates functions of the worldsheet coordinate $\sigma$.  The number of currents is double that of the original Lie algebra, since they are also worldsheet vectors.  (I.e., there are $\tau$ and $\sigma$ components, or ``left" and ``right", depending on the basis.  In the present case, the left and right currents are also left-propagating and right-propagating on the worldsheet; this is determined by the definition of the Virasoro operators, which we don't discuss here.)  The covariant derivatives and symmetry generators become currents $Z$ and $\tilde Z$ that commute with each other, $[Z,\tilde Z]=0$.
\item The zero-modes of this affine Lie algebra define an enlarged ordinary Lie algebra/group, the inhomogeneous version IG of the original group G \cite{Machiko}.  For manifest T-duality, double the coordinates to describe this enlarged group space, using the standard construction for the affine Lie algebra of a group \cite{Ed}.  
\item Make this group space into a general curved space (describing massless fields) by multiplying the covariant derivative currents $Z$ by a ``vielbein" $E$:  The group currents $Z$ are thus a basis for general currents $\Pi$ on this space; they define the ``vacuum", $\langle\Pi\rangle = Z$.  The algebra of these currents $\Pi$ replaces the structure constants of the affine Lie algebra IG with covariant ``torsion".  Requiring that the inhomogeneous term still gives the group metric imposes orthogonality on the vielbein.
\item The coset constraints are then imposed by requiring that commutators of the currents $\Pi$ of H with arbitrary $\Pi$'s yield the same result as in the coset (before introducing the vielbein).  This implies the $\Pi$ for H can be gauged to its coset value, and fixes the H-dependence of the remaining currents.  These constraints can be stated as conditions on the torsion.
\item Apply any additional torsion constraints, such as those in ordinary (super)gravity.
\item Finally, to spontaneously break T-duality symmetry and return to the usual coordinates, half of the currents for the symmetry generators $\tilde Z$ (forming a subalgebra) are taken as Killing vectors \cite{ran}.  (This corresponds to removing the coordinates for the inhomogeneous part of IG, reversing step 3 above.)  Since they commute with the basis $Z$ for the covariant derivatives, the requirement that they commute with the (curved space) covariant derivatives $\Pi$ implies that the vielbein $E$ is independent of the corresponding coordinates.
\end{enumerate}

\section{Coset spaces and their generalizations}

We briefly review coset spaces, their generalizations and related constructions like the covariant derivatives. For further information see \cite{1001}.

\subsection{Group spaces}

\label{pumpkaPreLenku}

Coset constructions have proven useful in defining representations of the Poincar\'e, (anti) de Sitter, and conformal groups, and their supersymmetric generalizations.  
With these in mind, we now review the general procedure for defining fields on coset spaces.

Cosets are often used to construct nonlinear $\sigma$ models:  There one focuses on the coset space itself, of which the scalar fields are elements.  For example,  one usually first-quantizes string theory about symmetric backgrounds by treating the spacetime coordinates $X(\tau,\sigma)$ (etc.)\ as coordinates of a coset space.  (Of course, more general backgrounds are also considered, but are less tractable.)  The string wave function is then implicitly a scalar functional of these coordinates (at fixed $\tau$).

There is some difficulty with this approach for the superstring, since the ground state, and thus the string field/wave function, is not a scalar.  Similar remarks apply to introducing massless backgrounds into the string action, since the coordinates carry ``curved" indices, while coupling gravity to fermions requires also ``flat" ones.

The generalization that solves this problem is simple:  For the coset G/H, keep all the coordinates of G (the ``symmetry" or ``isometry" group), rather than the usual procedure of immediately going to a unitary gauge where the coordinates of H (the ``gauge", ``isotropy", or ``stabilizer" subgroup) are gauged away.  The dependence of the fields on the H coordinates will be fixed, by defining their representations of H, but will be trivial only for scalars.

For this purpose we need to distinguish the differential operators responsible for left and right group multiplication:
\be
g' = g_L \, g \, g_R
\ee
Parametrizing any group element $g$ by coordinates $\alpha^I$ in terms of the generators $G_I$ 
\be
[\,G_{I},\,G_{J}\,]\,=\,-if_{IJ}{}^K\,G_{K}
\ee
(e.g., using any exponential parametrization), we can then write the corresponding infinitesimal transformations as
\be
\delta g \ = \ i\epsilon_L^I G_I g + g i\epsilon_R^I G_I = ( \epsilon_L^I q_I + \epsilon_R^I D_I ) g (\alpha)
\ee
where
\begin{equation}
q_{I}\,=\,L_{I}^{\,\,\,M}(\alpha)\partial_{M} , \quad
(dg)g^{-1}\,\equiv\,id\alpha^{M}L_{M}^{\,\,\,\,\,\,I}\,G_{I}
\end{equation}
\begin{equation}
D_{I}\,=\,R_{I}^{\,\,\,M} ( \alpha ) \partial_M , \quad
g^{-1}(dg)\,\equiv\,id\alpha^{M}R_{M}^{\,\,\,\,\,\,\,I}\,G_{I}
\end{equation}
(where $\partial_{M}\,\equiv\,{\partial}/{\partial\,\alpha^{M}}$) define the symmetry generators $q$ and covariant derivatives $D$ in terms of the vielbein appearing in the differential forms invariant under one or the other type of transformation.  Because left and right group multiplication commute, so do the symmetry generators and covariant derivatives:
\be
[ q_I , D_J ] \,=\, 0
\ee
Thus the ``covariant" derivatives are actually {\it in}variant; they become only covariant in unitary H gauges, due to compensating gauge transformations.

\subsection{Fields on coset spaces}
We decompose the basis of generators $G_I$ of the symmetry group G into the generators $H_\iota$ of the isotropy group H and the remaining ones $T_i$ of the coset G/H.
The representation space for the coset is constructed as follows: Define the linear space with basis elements $\Bra{0,\,_{m}}$. Let that space carry the matrix representation $\rho\,(H_{\iota})^{k}{}_{m}$ of the isotropy subgroup algebra; i.e., we have: 
\begin{equation}
\Bra{0,\,_{m}}\,H_{\iota}\,\defeq\,\Bra{0,\,_{k}}\,\rho\,(H_{\iota})^{k}{}_{m}
\end{equation}
We also have the action of the whole group on this basis: 
\be
\Bra{\alpha,\,_{m}}\,\defeq\,\Bra{0,\,_{m}}\,g^{-1}(\alpha)
\ee
We can then express the representation of the symmetry generators and covariant derivatives as differential operators on the wave function 
\be
\psi_{m}(\alpha)\,\defeq\,\Bra{\alpha,\,_{m}}\psi\,\rangle
\ee
The wave function $\psi_{m}\,(\alpha)$ depends also on the isotropy group coordinates $\alpha^{\iota}$, but this dependence is fixed:  In a convenient exponential parametrization, 
\begin{equation}
\label{newFlat}
\begin{array}{lllll}
\psi_{m}(\alpha)&\defeq&\Bra{0,\,_{m}}\,e^{-i\,\alpha^{\iota}\,H_{\iota}}\,e^{-i\,\alpha^{i}\,T_{i}}\,\Ket{\psi}&=&{\Big{(}}e^{-i\,\alpha^{\iota}\,\rho\,(\,H_{\iota}\,)}{\Big{)}}_{m}^{\,\,\,\,\,k}\,\Bra{0,\,_{k}}\,e^{-i\,\alpha^{i}\,T_{i}}\,\Ket{\psi}\\
&=&{\Big{(}}e^{-i\,\alpha^{\iota}\,\rho\,(\,H_{\iota}\,)}{\Big{)}}_{m}^{\,\,\,\,\,k}\,\psi_{k}\,(\alpha^{i})&\equiv&e_{m}{}^{k}(\alpha^{\iota}\,)\,\psi_{k}\,(\alpha^{i})\\
\end{array}
\end{equation}
The vielbein $e_{m}{}^{k}(\alpha^{\iota})$ is dependent only on the isotropy group coordinates $\alpha^{\iota}$ and can be gauged to the identity. 

From the above construction we know how the covariant derivatives corresponding to the isotropy subgroup act on $\psi_m(\alpha)$: 
\begin{equation}
\begin{array}{lll}
D_{\iota}\,\psi_{m}\,(\alpha)&=&\Bra{0,\,_{k}}\,-\,i\,\rho\,(H_{\iota})_m{}^k\,g^{-1}(\alpha)\,\Ket{\psi}\\
&=&-\,i\,\rho\,(H_{\iota}){}_m{}^k\psi_{k}\,(\alpha)\\
\end{array}
\end{equation}
We can also calculate the action of the symmetry group generators on the wave function: 
\begin{equation}
\begin{array}{lll}
q_{I}\,\psi_{m}(\alpha)&=&\Bra{0,\,_{m}}\,g^{-1}(\alpha)\,G_{I}\,\Ket{\psi}\\
&=&(\,G_{I}\,\psi\,)_{m}\,(\alpha)\\
\end{array}
\end{equation}

Since we know how the covariant derivatives with respect to the $\alpha^{\iota}$ act, we can therefore solve those constraints and replace partial derivatives (with respect to the $\alpha^\iota$) with matrices in $q_{I}$ and $D_{I}$. The dependence of all objects on the isotropy group coordinates is thus fixed. The remaining covariant derivatives $D_{i}$ act nontrivially.

\subsection{Curved spaces with isotropic coordinates}

\label{Curved}

We can also covariantize the covariant derivatives $D_{I}$ with respect to (super) Yang-Mills symmetry, see \cite{1001}. (The (super) Yang-Mills gauge group is unrelated to the isotropy gauge group, except for the case of gravity.) We can write the (super) Yang-Mills covariantized covariant derivatives as: 
\be
\nabla_{I}\,\defeq\,D_{I}\,+\,i\,A_{I} , \quad [\,\nabla_{I},\,\nabla_{J}\,\}\,=\,f_{IJ}{}^K\,\nabla_{K}\,+\,i\,F_{IJ}
\ee

In the first-quantized approach to (super)gravity the derivatives are gauge covariantized with respect to the (super-)Poincar\'{e} group  \cite{warren2}. The Yang-Mills generators are replaced with partial derivatives with respect to all coordinates:
\be
D_{I}\,\rightarrow\,\nabla_{I}\,=\,e_{I}^{\,\,\,\,K}\,\partial_{K}\,=\,\hat{e}_{I}^{\,\,\,\,K}\,D_{K}
\ee
The vielbein $e_{I}^{\,\,\,\,K}$ or  $\hat{e}_{I}^{\,\,\,\,K}$ are arbitrary.  The local Lorentz transformations are now included with the rest of the coordinate transformations and the covariant derivatives transform under the symmetry transformations as: 
\be
\nabla^{'}\,=\,e^{\Lambda}\,\nabla\,e^{-\Lambda} \quad\hbox{where}\quad \Lambda\,\defeq\,\Lambda^{M}\,D_{M}\,\equiv\,\bar{\Lambda}^{A}\,\nabla_{A}
\ee

The torsion $T$ is a combination of the structure constants and field strengths of Yang-Mills:
\begin{equation}
\label{dany}
[\,\nabla_{I},\,\nabla_{J}\,]\,=\,T_{IJ}{}^K\,\nabla_{K} 
\end{equation}
We divide indices as in the section 2, for the isotropy group, which in our case will be the Lorentz groups $SO(D-1,1)^{2}$, and for the coset space: We can write $\nabla_{I}\,\equiv\,(\,\nabla_{H},\,\nabla_{G/H}\,)$. Using the newly defined indices:
\begin{equation}
\label{trtko330}
\begin{array}{rcl}
[\,\nabla_{\,H},\,\nabla_{\,H}\,]&=&f_{\,H\,\,H}{}^{H}\,\nabla_{\,H}\\
{[}\nabla_{\,H},\,\nabla_{G/H}{]}&=&f_{\,H\,\,\,G/H}{}^{G/H}\,\nabla_{\,G/H}\\
{[}\nabla_{\,G/H},\,\nabla_{G/H}{]}&\equiv&R_{\,G/H\,\,\,G/H}{}^{H}\,\nabla_{\,H}\,+\,T_{\,G/H\,\,\,\,G/H}{}^{G/H}\,\nabla_{\,G/H}\\
\end{array}
\end{equation} 
The $R$ in (\ref{trtko330}) is the usual curvature (its stringy analog will be calculated in the Riemann tensor subsection \ref{Riemann}); $T_{\,G/H\,\,\,\,G/H}{}^{G/H}$ is the usual torsion. 

We have required that $\nabla_H$ act as in coset space (which in our case will be flat space):  The fact that the torsions $T_{H\,H}{}^{H}$ and $T_{\,H\,\,H}{}^{G/H}$ (=\,0) take their free values implies that $\nabla_H$ can be gauged to its free value.
The isotropy transformation of the coset part $\nabla_{G/H}$ is fully fixed by the requirement that the torsions $T_{H\,\,G/H}{}^{H}$ (=\,0) and $T_{H\,\,G/H}{}^{G/H}$ get their free values.  (We will see the stringy analog of this in subsection \ref{srdieckoPreLenku}.)
By keeping this dependence on the H coordinates, rather than gauging them away entirely, we have the first-quantized way to define the spin (for arbitrary representations), as a differential operator on that space, see \cite{warren2}.     

\section{Affine Lie algebra and generalized T-duality}

\subsection{Current algebras}

For application to the string, we consider current algebras on the worldsheet, or affine Lie algebras
\begin{equation}
\mbox{[}\,Z_{\mbox{{\scriptsize\Fontamici{M}}}}\,\xA,\,Z_{\mbox{{\scriptsize\Fontamici{N}}}}\,\xB\,\mbox{]}\,=\,-i\,\eta_{\mbox{{\scriptsize\Fontamici{M}}}\,\mbox{{\scriptsize\Fontamici{N}}}}\,\deltP\,-\,i\,f_{\mbox{{\scriptsize\Fontamici{M}}}\,\mbox{{\scriptsize\Fontamici{N}}}}{}^{\mbox{{\scriptsize\Fontamici{P}}}}\,Z_{\,\mbox{{\scriptsize\Fontamici{P}}}}\,\delt
\label{twotwo}
\end{equation} 
where $f$ is the structure constants of the ordinary Lie algebra.  (Note that all the generators are understood as string currents, so they are dependent on the string coordinate $\sigma\,\equiv\,\sigma_{1}\,\equiv$ ``1".
There is an implicit $2\pi$ with every $\delta(\sigma)$.  Also, for dimensional analysis there is an implicit $1/\alpha'$ with $\eta$.)  The metric $\eta$ of the affine (Schwinger) term is invertible as a consequence of our including both components of the current, as should be clear from the Abelian case considered below.
Due to our doubling of coordinates for manifest T-duality, 
the group coordinates $X^{\mbox{{\scriptsize\Fontamici{M}}}\,}$ carry the same index.
Acting on background fields $\phi$, these currents reduce to 
the group covariant derivatives $D_{\mbox{{\scriptsize\Fontamici{M}}}}$ of the ordinary (non-affine) algebra
(with the same structure constants),
\be
[ Z_{\go M} \xA , \phi ( X \xB ) ] \,=\, -i (D_{\go M} \phi ) \delt
\ee
(Similar remarks apply to a second Lie algebra $\tilde Z$ for which $q$ replaces $D$ and $[Z,\tilde Z]=0$.)
We are interested in the affine Poincar\'e algebra,
where the index
\be
_{\mbox{{\scriptsize\Fontamici{M}}}}\,\defeq\,(\,_{MN},\,_{M},\,^{MN}\,)
\ee
has dimension $2D^{2}$, as we will now describe.

We begin with the current algebra associated with the usual $X$ coordinates.
In string theory one naturally gets the interpretation of T-duality as the reflection subgroup of the bigger $O(D,D)$ group. One can rewrite the string oscillator algebra using the explicit $O(D,D)$ vector 
\be
P_{M}\,\defeq\,(\,P_{m},\,X'^{m}\,)
\ee
Using this generalized $O(D,D)$ momentum one gets the algebra 
\begin{equation}
[\,P_{M}\,(\mbox{{\footnotesize{1}}}),\,P_{N}\,(\mbox{{\footnotesize{2}}})\,]\,=\,-\,i\,\eta_{M\,N}\,\deltP
\label{dd}
\end{equation}
where $\eta_{MN}$ is the $O(D,D)$ metric:
\be
\eta_{MN}\,=\,
\begin{pmatrix}
{ 0}& { \delta_{m}^{\,\,\,n}}\\
 { \delta_{n}^{\,\,\,m}} &{ 0}\\
\end{pmatrix}
\ee

In the future we want to use a different basis for the string oscillator algebra (\ref{dd}). Therefore we introduce the left/right vector 
\be
\label{LR}
P_{M}\,\defeq\,(\,P_{\bf{m}},\,P_{\tilde{\bf{m}}}\,)\,\equiv\,\frac{1}{\sqrt{2}}\,(\,P_{m}\,+\,X^{'}_{m},\,P_{m}\,-\,X^{'}_{m})
\ee
In this basis the oscillator algebra has the same form as (\ref{dd}) except for the form of the metric:
\be
\eta_{MN}\,=\,
\begin{pmatrix}
{ \eta_{\bf{m\,n}}}& { 0}\\
 { 0} &{ -\,\eta_{\bf{\tilde{m}\,\tilde{n}}}}\\
\end{pmatrix}
\ee

\subsection{Lorentz}

In the next step we want to merge the algebra (\ref{dd}) with the Lorentz algebra $so(D-1,1)^{2}$.  The reason is that the metric $g$ and $b$ field are in the coset space $SO(D,D)/SO(D-1,1)^{2}$. This suggests that the coordinate space should be obtained by modding out by the subgroup \mbox{$SO(D-1,1)^{2}$}. The left/right basis of (\ref{LR}) is then appropriate. 

The generators for this Lorentz algebra are denoted as
\be
S_{MN}\,\defeq\,(\,S_{\bf mn},\,S_{\bf\tilde m\tilde n}\,)
\ee
and satisfy the usual commutation relations
\begin{eqnarray}
\label{trtkoMinus}
[\,S_{\bf mn}\,{\mbox{\footnotesize(1)}},\,S_{\bf kl}\,{\mbox{\footnotesize(2)}}\,]&=&i\eta_{[\,\bf{m}\,[\,\bf{k}}\,S_{\bf{n}\,]\,\bf{l}\,]}\,\,\delta\,(\,{\mbox{\footnotesize{2}}}\,-\,{\mbox{\footnotesize 1}}\,)\\
\mbox{[}\,S_{\bf{m}\,\bf{n}}\,{\mbox{\footnotesize(1)}},\,S_{\tilde{\bf{k}}\,\tilde{\bf{l}}}\,{\mbox{\footnotesize(2)}}\,\mbox{]}&=&0\\
&{\xymatrix{\ar@/_/@{>}[r]&}}&\mbox{\footnotesize{Same for Left $\,\rightarrow\,$ Right}}\nonumber
\end{eqnarray}
(where $[\dots]$ is the unweighted anti-symme\-trization).  Since $P$ and $S$ form the ordinary Poincar\'{e} algebra, we have: 
\begin{eqnarray}
\label{trtkoPlus}
\mbox{[}\,S_{\bf{m}\,\bf{n}}\,\xA,\,P_{\bf{k}}\,\xB\,\mbox{]}&=&i\eta_{\bf{k}\,[\,\bf{m}}\,P_{\bf{n}\,]}\,\delt \\
\mbox{[}\,S_{\bf{m}\,\bf{n}}\,\xA,\,P_{\tilde{\bf{k}}}\,\xB\,\mbox{]}&=&0\\
&{\xymatrix{\ar@/_/@{>}[r]&}}&\mbox{\footnotesize{Same for Left $\,\rightarrow\,$ Right}}\nonumber
\end{eqnarray}

However, the set of generators $(S_{MN},\,P_{M})$ does not form a closed affine Lie algebra. The Jacobi identity requires a new field $\Sigma$ such that 
\be
[\,P,\,P\,]\,\propto\,\delta^{'}\,+\,\Sigma\quad \hbox{and}\quad[\,S,\,\Sigma\,]\,\propto\,\delta^{'}\,+\,\Sigma
\ee
Using the commutators $[\,S,\,[\,P,\,P\,]\,]$ and the Jacobi identity, we obtain the new set of generators 
\be
Z_{\mbox{{\scriptsize\Fontamici{M}}}}\,\defeq\,(\,S_{M\,N},\,P_{M},\,\Sigma^{M\,N}\,)
\ee
for which we have the following affine Lie algebra: 
\begin{eqnarray}
\label{algebra}
{[}\,S_{\bf{m}\,\bf{n}}\,{\mbox{\footnotesize(1)}},\,S_{\bf{k}\,\bf{l}}\,{\mbox{\footnotesize(2)}}\,{]}&=
&-i\eta_{[\,\bf{m}\,[\,\bf{k}}\,S_{\bf{l}\,]\,\bf{n}\,]}\,\,\delta\,(\,{\mbox{\footnotesize{2}}}\,-\,{\mbox{\footnotesize 1}}\,)\\
{[}\,S_{\bf{m}\,\bf{n}}\,\xA,\,P_{\bf{k}}\,\xB\,{]}&=&i\eta_{\bf{k}\,[\,\bf{m}}\,P_{\bf{n}\,]}\,\delt \nonumber\\
{[}\,S_{\bf{m}\,\bf{n}}\,\xA,\,\Sigma^{\bf{k}\,\bf{l}}\,\xB\,{]}&=
&-i\,\delta_{\bf mn}{}^{\bf kl}\,\deltP\,-\,i\delta_{[\,\bf{m}}{}^{[\,\bf{k}}\,\eta_{\bf{n}\,]\,\bf{s}}\Sigma^{\bf{l}]\bf{s}}\,\delt\nonumber\\
{[}\,P_{\bf{m}}\,\xA,\,P_{\bf{n}}\,\xB\,{]}&=
&-\,i\,\eta_{\bf mn}\,\deltP\,+\,i\eta_{\bf mh}\,\eta_{\bf{n}\,\bf{s}}\Sigma^{\bf hs}\,\delt\nonumber\\
{[}\,P_{\bf{m}}\,\xA,\,\Sigma^{\bf{k}\,\bf{l}}\,\xB\,{]}&=&0\nonumber\\
{[}\,\Sigma^{\bf{m}\,\bf{n}}\,\xA,\,\Sigma^{\bf{k}\,\bf{l}}\,\xB\,{]}&=&0\nonumber\\
&{\xymatrix{\ar@/_/@{>}[r]&}}&\mbox{\footnotesize{Same for Left $\,\rightarrow\,$ Right}}\nonumber\\
{[}\,\mbox{Left},\,\mbox{Right}\,{]}\,&=&\,0\nonumber
\end{eqnarray}
Thus we get the general structure of an affine Lie algebra (\ref{twotwo}).  (Non-affine stringy Lorentz algebras were considered in \cite{Erg}.  Left and right spin algebras have also been used in \cite{Nat}, but commuting with $P$.  Neither of those had $\Sigma$.) 

For dealing with antisymmetric pairs of indices we have introduced an implicit metric such that for any two antisymmetric tensors we have
\be
A\cdot B \,\equiv\, \frac12 A^{\bf mn}B_{\bf mn}
\ee
The identity matrix with respect to this inner product is
\be
\delta_{\bf mn}{}^{\bf pq} \,\equiv\, \delta_{[\bf m}{}^{\bf p} \delta_{{\bf n}]}{}^{\bf q}
\ee

The only nonvanishing terms in the metric and structure constants are (as could be guessed by dimensional analysis)
\be
\eta_{PP} , \ \eta_{S\Sigma}\, ; \quad f_{SPP} , \  f_{SS\Sigma}
\ee
where we have lowered the upper index on $f$ with $\eta$ to take advantage of its total antisymmetry, and used ``schematic" notation, replacing explicit indices with their type:
\be
\mbox{{\scriptsize\Fontamici{M}}}
\,\defeq\,(\,_{MN},\,_M,\,^{\,MN\,}\,)
\,\defeq\,(\,S,\,P,\,\Sigma\,)
\ee
Explicitly these are, for the left-handed algebra,
\be
(\eta)_{\bf mn} \,=\, \eta_{\bf mn} , \  (\eta)_{\bf mn}{}^{\bf pq} \,=\, \delta_{\bf mn}{}^{\bf pq} \, ; \quad
f_{\bf mn}{}^{\bf pq} \,=\, - \delta_{\bf mn}{}^{\bf pq} , \  
f_{\bf mn\,pq}{}^{\bf rs} \,=\, \eta_{\bf [m[p} \delta_{\bf q]n]}{}^{\bf rs}
\ee
For the right-handed algebra we change the signs of the corresponding terms in $\eta_{\go{MN}}$ but not in $f$.

\section{Curved spaces with affine algebras}

\subsection{Background fields}

We now introduce background fields following \cite{warren}, but using the affine algebra (\ref{twotwo}). Using the vielbein we can write: 
\be
\Pi_{\mbox{{\scriptsize\Fontamici{A}}}}\xA\,=\,E_{\mbox{{\scriptsize\Fontamici{A}}}}{}^{{\mbox{{\scriptsize\Fontamici{M}}}}}(X^{\mbox{{\scriptsize\Fontamici{M}}}})Z_{\mbox{{\scriptsize\Fontamici{M}}}}
\ee
Then we get the affine Lie algebra for the $\Pi_{\mbox{{\scriptsize\Fontamici{A}}}}$ operators:
\begin{equation}
\begin{array}{cccccc}
\label{Lenuska1}
\mbox{[}\Pi_{\mbox{{\scriptsize\Fontamici{A}}}}\xA,\Pi_{\mbox{{\scriptsize\Fontamici{C}}}}\xB\mbox{]}\,\equiv\,-i\eta_{\mbox{{\scriptsize\Fontamici{A}}}\mbox{{\scriptsize\Fontamici{C}}}}\,\deltP-iT_{\mbox{{\scriptsize\Fontamici{A}}}\mbox{{\scriptsize\Fontamici{C}}}}{}^{\mbox{{\scriptsize\Fontamici{E}}}}\Pi_{\mbox{{\scriptsize\Fontamici{E}}}}\,\delt
\end{array}
\end{equation}
where $T$ is the stringy generalization of the torsion:
\be
\label{torsion}
T_{\mbox{{\scriptsize\Fontamici{A}}}\mbox{{\scriptsize\Fontamici{C}}}}{}^{\mbox{{\scriptsize\Fontamici{E}}}}=E_{[\mbox{{\scriptsize\Fontamici{A}}}}{}^{\mbox{{\scriptsize\Fontamici{M}}}}(D_{\mbox{{\scriptsize\Fontamici{M}}}}E_{\mbox{{\scriptsize\Fontamici{C}}}]}{}^{\mbox{{\scriptsize\Fontamici{N}}}})E^{-1}_{\mbox{{\scriptsize\Fontamici{N}}}}{}^{\mbox{{\scriptsize\Fontamici{E}}}}
+\frac{1}{2}\eta^{\mbox{{\scriptsize\Fontamici{E}}}\mbox{{\scriptsize\Fontamici{D}}}}E_{\mbox{{\scriptsize\Fontamici{D}}}}{}^{\mbox{{\scriptsize\Fontamici{M}}}}(D_{\mbox{{\scriptsize\Fontamici{M}}}}E_{[\mbox{{\scriptsize\Fontamici{A}}}|}{}^{\mbox{{\scriptsize\Fontamici{N}}}})E^{-1}_{\mbox{{\scriptsize\Fontamici{N}}}}{}^{\mbox{{\scriptsize\Fontamici{F}}}}\eta_{\mbox{{\scriptsize\Fontamici{F}}}|\mbox{{\scriptsize\Fontamici{C}}}]}
+E_{\mbox{{\scriptsize\Fontamici{A}}}}{}^{\mbox{{\scriptsize\Fontamici{M}}}}E_{\mbox{{\scriptsize\Fontamici{C}}}}{}^{\mbox{{\scriptsize\Fontamici{N}}}}E^{-1}_{\mbox{{\scriptsize\Fontamici{P}}}}{}^{\mbox{{\scriptsize\Fontamici{E}}}}f_{\mbox{{\scriptsize\Fontamici{M}}}\mbox{{\scriptsize\Fontamici{N}}}}{}^{\mbox{{\scriptsize\Fontamici{P}}}}
\ee
where $[\,\go A\,|\,|\,\go C\,]$ indicates antisymmetrization in only those indices.
Note that the Jacobi identities imply the total antisymmetry of the torsion, just as for the structure constants.

This torsion can be identified with that of ``ordinary" curved-space covariant derivatives (as in subsection \ref{Curved}) by use of the strong constraint:  We write
\be
\nabla_{\mbox{{\scriptsize\Fontamici{A}}}}\,\defeq\,E_{\mbox{{\scriptsize\Fontamici{A}}}}{}^{\mbox{{\scriptsize\Fontamici{M}}}}D_{\mbox{{\scriptsize\Fontamici{M}}}}
\ee
Using this and the strong constraint
\be
(\nabla^{\mbox{{\scriptsize\Fontamici{A}}}}\phi)(\nabla_{\mbox{{\scriptsize\Fontamici{A}}}}\psi)\,=\,0
\ee
we get the same torsion in
\begin{equation}
\label{trtko29}
[\,\nabla_{\mbox{{\scriptsize\Fontamici{A}}}},\,\nabla_{\mbox{{\scriptsize\Fontamici{C}}}}\,] = \,T_{\mbox{{\scriptsize\Fontamici{A}}}\,\mbox{{\scriptsize\Fontamici{C}}}}{}^{\mbox{{\scriptsize\Fontamici{D}}}}\,\nabla_{\mbox{{\scriptsize\Fontamici{D}}}}
\end{equation}
when acting on fields, since the second term in (\ref{torsion}) can be added for free.

By setting the coefficient of the Schwinger term to be the metric $\eta$, the vielbein is forced to obey the orthogonality constraints: 
\begin{equation}
E_{\mbox{{\scriptsize\Fontamici{A}}}}{}^{\mbox{{\scriptsize\Fontamici{M}}}}\eta_{\mbox{{\scriptsize\Fontamici{M}}}\mbox{{\scriptsize\Fontamici{N}}}}\,E_{\,\mbox{{\scriptsize\Fontamici{C}}}}{}^{\mbox{{\scriptsize\Fontamici{N}}}}\,\equiv\,\eta_{\mbox{{\scriptsize\Fontamici{A}}}\mbox{{\scriptsize\Fontamici{C}}}}
\end{equation}
This choice  does not affect the physics, and simplifies many of the expressions.  For example, it implies the total antisymmetry of the torsion, when the upper index is implicitly lowered with $\eta$:
\be
T_{\go{A\,B\,C}} \,=\,\frac12 E_{[\,\go A\,|}{}^{\go M}(D_{\go M} E_{|\,\go B}{}^{\go N})E_{\go C\,]\,\go N}
	+ E_{\go A}{}^{\go M} E_{\go B}{}^{\go N} E_{\go C}{}^{\go P} f_{\go{M\,N\,P}}
\ee
where we have used $E^{-1}_{\,\,\,\mbox{{\scriptsize\Fontamici{M}}}}{}^{\,\mbox{{\scriptsize\Fontamici{A}}}}\,=\,\eta^{\mbox{{\scriptsize\Fontamici{A\ B}}}}\eta_{\mbox{{\scriptsize\Fontamici{M\ N}}}}E_{\,\mbox{{\scriptsize\Fontamici{B}}}}{}^{\mbox{{\scriptsize\Fontamici{N}}}}$.  (Also note that in the first term the antisymmetrization can be written as a cyclic sum without the 1/2, since it is already antisymmetric in the last two indices.)  Thus, because of orthogonality, the vielbein is like (the exponential of) a 2-form, while the torsion is a 3-form; similarly, the Bianchi identities are a 4-form.

When solving the orthogonality constraint, note that we are also putting some parts of $E$ to zero or to some particular constant value, which comes from the coset constraints on the torsion, as explained later. We get:
\begin{equation}
\label{B1}
E_{\mbox{{\scriptsize\Fontamici{A}}}}{}^{\mbox{{\scriptsize\Fontamici{M}}}}\,=\ 
\bordermatrix{
& {}^{MN} & {}^M & {}_{MN} \cr
{}_{AB} & \delta_{AB}{}^{MN}& 0 & 0 \cr
{}_A & \omega_A{}^{MN} & e_A{}^M & 0 \cr
{}^{AB} & r^{ABMN}\,-\,\frac{1}{2}\,\omega^{CAB}\,\omega_C{}^{MN} & -e_C{}^M\omega^{CAB} & \delta^{AB}{}_{MN} \cr}
\end{equation}
where the new fields $e$, $\omega$ and $r$ were introduced. The $r$ has a role to be explained later, and satisfies
\begin{equation}
r^{ABCD}\,+\,r^{CDAB}\,=\,0
\end{equation}
 
\subsection{Coset constraints}

\label{srdieckoPreLenku}

Our aim is to generalize the coset construction described in subsection \ref{Curved} to affine Lie algebras, specifically the affine Poincar{\'e} algebra (\ref{algebra}).  Isotropy group dependence is fixed by the constraint that the covariant derivatives with the Lorentz group indices $S\,\equiv\,_{A\,B}$ act on fields by some particular matrix representation, i.e., 
\be
(\nabla_{S}\,\psi)_{\,S\,}\,\defeq\,(M_S)_S{}^S\,\psi_{\,S\,}
\ee
For the covariant derivatives themselves, this implies, as described in section \ref{Curved},
\be
\label{coset}
[ \nabla_S , \nabla_{\mbox{{\scriptsize\Fontamici{A}}}\ } ] = f_{S\mbox{{\scriptsize\Fontamici{A}}}}{}^{\mbox{{\scriptsize\Fontamici{B}}} }\nabla_{\mbox{{\scriptsize\Fontamici{B}}}}
\qquad (  T_{S\mbox{{\scriptsize\Fontamici{A}}}}{}^{\mbox{{\scriptsize\Fontamici{B}}} }  = f_{S\mbox{{\scriptsize\Fontamici{A}}}}{}^{\mbox{{\scriptsize\Fontamici{B}}}\,} )
\ee
I.e., all covariant derivatives are in the same representations of $S$ as in flat space.  In particular, this means the subalgebra of $\nabla_S$ is unmodified from flat space, so we can choose the gauge
\be
\nabla_S = D_S
\qquad ( E_S{}^{\mbox{{\scriptsize\Fontamici{M}}}} = \delta_S{}^{\mbox{{\scriptsize\Fontamici{M}}}\,} )
\ee
(However, other gauges, such as lightcone gauges, may also be useful \cite{warren2}.)  This gauge was used, in addition to orthogonality, to obtain the expression for the vielbein in (\ref{B1}). 

The rest of the coset constraint (\ref{coset}) gives the action of $D_{S}$ on the nontrivial components of $E_{A}{}^{\go K}$:
\begin{equation}
\label{trtko31}
\begin{array}{lllll}
D_S\,E_P{}^P&\equiv&D_{AB}\,e_C{}^K&=
&-\eta_{C[A}\,e_{B\,]}^{\,\,\,\,K}\,+\,\,e_{C}{}^M\eta_{M[A}\,\delta_{B]}{}^K\,\\
D_{S}\,E_{P}{}^{S}&\equiv&D_{AB}\,\omega_{\,C}{}^{KL}&=
&-\,\eta_{C[A}\,\omega_{B\,]}{}^{KL}\,+\,\omega_{C}{}^{MN}\,\eta_{[M[\,A}\,\delta_{B]}{}^{K}\,\delta_{N]}{}^{L}\,\\
\end{array}
\end{equation}
Thus in this gauge the dependence on the Lorentz coordinates is fixed for the vielbein, as well as the (residual) gauge parameters.  (E.g., the Lorentz gauge parameters still have arbitrary dependence on $x$.)

Dimensional analysis is useful for further analysis of the torsion. The following table summarizes the torsion engineering dimensions:

\begin{center}
\begin{tabular}{| cc |}
\noalign{\vspace{.1in}}
\hline
\texttt{Torsion component}&\texttt{Dimension}\\
\hline
$T_{S\,S}^{\,\,\,\,\,\,\,\,\,\Sigma}$&$-\,2$\\
\hline
$T_{S\,S}^{\,\,\,\,\,\,\,\,\,P}$&$-\,1$\\
\hline
$T_{S\,S}^{\,\,\,\,\,\,\,\,\,S}$&$0$\\
\hline
$T_{S\,P}^{\,\,\,\,\,\,\,\,\,P}$&$0$\\
\hline
$T_{S\,P}^{\,\,\,\,\,\,\,\,\,S}$&$1$\\
\hline
\noalign{\vspace{.1in}}
\end{tabular}
\quad
\begin{tabular}{| cc |}
\noalign{\vspace{.1in}}
\hline
\texttt{Torsion component}&\texttt{Dimension}\\
\hline
$T_{P\,P}^{\,\,\,\,\,\,\,\,\,P}$&$1$\\
\hline
$T_{S\,\Sigma}^{\,\,\,\,\,\,\,\,\,S}$&$2$\\
\hline
$T_{P\,P}^{\,\,\,\,\,\,\,\,\,S}$&$2$\\
\hline
$T_{P\,\Sigma}^{\,\,\,\,\,\,\,\,\,S}$&$3$\\
\hline
$T_{\Sigma\,\Sigma}^{\,\,\,\,\,\,\,\,\,S}$&$4$\\
\hline
\noalign{\vspace{.1in}}
\end{tabular}
\end{center}
Note that most of the torsions, including all torsions of nonpositive dimension, have already been fixed by the coset constraint.

\section{Relations to previous tensors}

\subsection{Remaining torsion constraint}

The ``usual" torsion constraint (generalized to 2D-valued indices)
\begin{equation}
\label{trtko30}
T_{PP}{}^P\,=\,0
\end{equation}
eliminates the last surviving torsion of dimension 1, and gives the constraints that were previously found in \cite{warren} by a different method.  This can be expanded in schematic notation as
\begin{equation}
\label{trtko45}
0\,=\,T_{{\color{red}P}{\color{blue}P}{\color{Green}P}}
\,=\,\frac12 E_{[\rP|}{}^{{\go K}}(D_{\go K} E_{|\bP}{}^{{\go H}})E_{\gP]{\go H}}\,+\,E_{\rP}{}^{\go K} E_{\bP}{}^{\go H} E_{\gP}{}^{\go L} f_{{\go K}{\go H}{\go L}}
\end{equation}
(Colored indices are not summed.)

For comparison, the analog of the torsion that appears in \cite{warren} (but taking into account orthogonality):
\begin{equation}
\label{trtko100}
{\bf F}_{ABC}\,\defeq\,\frac12 e_{[A|}{}^K(\partial_K e_{|B}{}^H)e_{C]H} 
\end{equation}
is the same except that the range of indices is over only $P$, where (in our gauge) $e_A{}^M\,\equiv\,E_A{}^M$ and $D_P=\partial_M$ acting on a field.
Thus, expanding the indices in (\ref{trtko45}) over $(S,\,P,\,\Sigma)$ will separate it into {\bf F} and $\omega$ terms.

Using the structure of the vielbein $E_{\,\mbox{{\scriptsize\Fontamici{A}}}}{}^{\mbox{{\scriptsize\Fontamici{M}}}}$ in (\ref{B1}), 
from the former term of (\ref{trtko45}) we get: 
\be
\hbox{\Cwarren}\,+\frac12 E_{[\rP|}{}^S(D_S E_{|\bP}{}^P)E_{\gP]P}
\ee
(Repeated schematic indices $(S,\,P,\,\Sigma)$ are summed over the subset indicated.)
The latter term in this expression vanishes according to the first condition in (\ref{trtko31}) and structure of the vielbein. The latter term of (\ref{trtko45}) gives: 
\begin{equation}
\label{trtko77}
\begin{array}{rcl}
\,E_{\rP}{}^{\go K} E_{\bP}{}^{\go H} E_{\gP}{}^{\go L} f_{{\go K}{\go H}{\go L}}
&=& \frac12 E_{[\rP}{}^S E_{\bP}{}^P E_{\gP]}{}^P\,f_{SPP}\\
&&\rP\rightarrow\,\rA\,\,\,|\,\,\,\bP\rightarrow\,\bB\,\,\,|\,\,\,\gP\rightarrow\,\gC\\
&=&\frac12 \omega_{[\rA\bB\gC]}
\end{array}
\end{equation}
We thus get the relation
\begin{equation}
\label{trtko1}
{\bf F}_{ABC} +\frac12 \omega_{[ABC]} \,=\, 0
\end{equation}
This agrees with the constraints on $\omega_{A}{}^{BC}$ in \cite{warren},
\be
\omega_{[\,{\bf abc}\,]}  \,=\, - 2 {\bf{F}}_{\bf abc}\,, \quad 
\omega_{{\bf a\tilde{b}\tilde{c}}} \,=\,-\, {\bf{F}}_{\bf a\tilde{b}\tilde{c}}
\ee

There are also constraints involving the dilaton, which work the same way as previously; these are needed to allow definition of a Ricci tensor and scalar (i.e., field equations and action) independent of those connections that are not fixed by the above constraint.

\subsection{Riemann tensor}

\label{Riemann}

Previously no full curvature tensor with manifest T-duality was derived, and even those pieces that were found came in an indirect way, not by commutation of covariant derivatives.  Here we duplicate the known curvature directly as a torsion, and the missing pieces are identified as corresponding to the new field $r^{ABCD}$.

From (\ref{trtko330}) the curvature tensor is $T_{\rP\,\bP}{}^{\gS}\,\equiv\,R_{\,{\color{red}{G/H}}\,\,\,{\color{blue}{G/H}}}{}^{{\color{Green}{H}}}$:
\begin{equation}
\label{trtko22}
\begin{array}{rcl}
T_{\rP\,\bP}{}^{\gS}&=&E_{[\,\rP}{}^{S}(D_{S}\,E_{\bP\,]}{}^{{\go R}})E^{-1}_{{\go R}}{}^{\,\gS}\,+\,\frac{1}{2}\,\eta^{\gS\,\Sigma}\,E_{\Sigma}^{\,\,\,\,S}(D_{S}\,E_{[\,\rP\,}^{\,\,\,\,\,\,\,{\go R}})E^{-1}_{{\go R}}{}^{\,{\go K}}\,\eta_{{\go K}\,|\,\bP\,]}\\
&+&E_{[\,\rP}^{\,\,\,\,\,\,\,\,P}(D_{P}\,E_{\bP\,]}{}^{{\go R}})E^{-1}_{{\go R}}{}^{\,\gS}\,+\,\frac{1}{2}\,\eta^{\gS\,\Sigma}\,E_{\Sigma}^{\,\,\,\,P}(D_{P}\,E_{[\,\rP\,}{}^{{\go R}})E^{-1}_{{\go R}}{}^{\,{\go K}}\,\eta_{{\go K}\,|\,\bP\,]}\\
&+&\frac{1}{2}\,\eta^{\gS\,\Sigma}\,E_{\Sigma}{}^{\Sigma}(D_{\Sigma}\,E_{[\,\rP\,}{}^{{\go R}})E^{-1}_{{\go R}}{}^{\,{\go K}}\,\eta_{{\go K}\,|\,\bP\,]}\\
&+&\,E_{\rP\,}{}^{S}E_{\bP\,}{}^{S}E^{-1}_{\,\,\,\,S}{}^{\,\,\gS}\,f_{S\,S}{}^{S}\,+\,E_{\,[\,\rP\,}{}^{S}\,E_{\bP\,]\,}{}^{P}\,E^{-1}_{\,\,\,\,P}{}^{\,\,\gS}\,f_{S\,P}{}^{P}
\,+\,E_{\rP\,}{}^{P}E_{\bP\,}{}^{P}E^{-1}_{\,\,\,\,\Sigma}{}^{\,\,\gS}\,f_{P\,P}{}^{\Sigma}
\end{array}
\end{equation}
Rewriting using explicit forms of the schematic indices and $f$,
and using (\ref{trtko31}) and (\ref{trtko100}),
after some algebra we get the final expression:
\begin{equation}
\label{trtkoFinal}
\begin{array}{rl}
T_{AB}{}^{CD}&=\ e_{[A}{}^M\,\partial_{\,M}\,\omega_{\,B\,]}{}^{CD}\,+\,\omega_{[\,A\,|}{}^C{}_{H}\,\omega_{B]}{}^{HD}\,-\,\frac{1}{2}\,\omega_{\,M}{}^{CD}\,\omega^M{}_{AB}\,-\,{\bf{F}}_{A\,B}{}^N\,\omega_N{}^{CD}\\
&+\,r^{C\,D}{}_{AB}\,+\,((D_\Sigma)^{CD}\,e_A{}^K)e_{BK}\\
\end{array}
\end{equation}

In the usual representations, $D_\Sigma=q_\Sigma=\partial_\Sigma$; as part of dimensional reduction, we set $q_\Sigma\,\phi\,=\,0$.  Then the curvature reduces to:
\begin{equation}
\label{trtkoFinal1}
T_{AB}{}^{C\,D}\,=\,e_{[\,A}^{\,\,\,\,\,\,M}\,\partial_{M}\,\omega_{B\,]}^{\,\,\,\,\,C\,D}\,+\,\omega_{[\,A\,|}^{\,\,\,\,\,\,\,C}{}_{H}\,\omega_{B\,]}^{\,\,\,\,H\,D}\,-\,\frac{1}{2}\,\omega_{M}^{\,\,\,\,\,\,\,C\,D}\,\omega^{M}_{\,\,\,\,A\,B}\,-\,{\bf{F}}_{A\,B}^{\,\,\,\,\,\,\,\,\,\,N}\,\omega_{N}^{\,\,\,\,\,C\,D}\,+\,r^{C\,D}_{\,\,\,\,\,\,\,\,\,\,A\,B}
\end{equation}
This form was derived also in \cite{warren} up to the antisymmetric $r^{CD}{}_{AB}$ part, required for covariance.  Here the curvature tensor was obtained in a more direct way. 

$r$ can also be fixed by constraining the corresponding part of the curvature to vanish:
\be
T_{\bf abcd} - T_{\bf cdab} \,=\, T_{\bf \tilde a\tilde b\tilde c\tilde d} - T_{\bf \tilde c\tilde d\tilde a\tilde b} \,=\, T_{\bf ab\tilde c\tilde d} - T_{\bf \tilde c\tilde dab} \,=\, 0
\ee

As the final step we reduce the coordinates to the usual half by dimensional reduction, with the conditions
\be
q_{\,\Sigma}\,\phi\,=\, (q_{P_L}-q_{P_{R}})\,\phi\,=\,0
\ee
Here $q$ indicates a Killing vector of the original (``flat") coset space, commuting with all the flat covariant derivatives $D$. Since $q_\Sigma$ are Abelian, we can always choose coordinates where $q_\Sigma=\partial_\Sigma$; and since the rest are Abelian mod $q_\Sigma$, we can also choose coordinates where they are $\partial_{P_L}-\partial_{P_R}$ mod $\partial_\Sigma$ terms.  We have also fixed the dependence of the fields on the Lorentz coordinates previously by the coset constraints. In that way the original $2D^2$-dimensional coordinate space is reduced to $\mathbb{R}^{D}$.

\section{Conclusion}

We outline the results we have obtained:
We began with the generalized affine algebra $SP\Sigma$ (\ref{algebra}), enlarging the configuration space to $2D^2$ dimensions. The background fields were introduced via the vielbein $E_{\mbox{{\scriptsize\Fontamici{A}}}}{}^{{\mbox{{\scriptsize\Fontamici{M}}}}}(X^{\mbox{{\scriptsize\Fontamici{N}}}})$. The orthogonality constraints were applied to them. Together with coset constraints on torsions the specific structure of the vielbein  was derived (\ref{B1}). From dimensional arguments we obtained one particular torsion constraint  reproducing that originally obtained in\nobreak\ \cite{warren}. From the torsion $T_{\rP\bP}{}^{\gS}\,\equiv\,R_{\,{\color{red}{G/H}}\,\,\,{\color{blue}{G/H}}}{}^{{\color{Green}{H}}}$ we got the curvature tensor. The result (\ref{trtkoFinal1}) matches the result from \cite{warren} except for the antisymmetric part $r^{CD}{}_{AB}$, which can be fixed by an additional constraint. The resulting curvature tensor has explicit $SO(D,D)$ index structure, which was our goal. 

Various generalizations suggest themselves:
\vspace{-8pt}
\begin{enumerate}
\setlength{\itemsep}{-4pt}
\item supersymmetry (especially AdS),
\item $\alpha'$ corrections, which may clarify the results of \cite{bow},
\item the corresponding first-quantization of the string (ghosts, BRST, etc.), and
\item string field theory (with vielbein fields).
\end{enumerate}

\section*{Acknowledgment}

This work was supported in part by National Science Foundation Grant No. PHY-0969739.

\end{document}